\documentclass[12pt]{iopart}
\usepackage{epsfig}
\begin{document}

\centerline{4th NuFact Workshop (Neutrino Factories based on Muon 
Storage Rings), London, UK, 1-6 Jul 2002}

\title{Muon acceleration with a very fast ramping synchrotron for a neutrino 
factory}

\author{D J Summers{$\,^1$}, J S Berg{$\,^2$}, A A Garren{$\,^3$} and 
R B Palmer{$\,^2$}  
\footnote[3]{To
whom correspondence should be addressed (summers@relativity.phy.olemiss.edu)}
}

\address{$^1$\ Dept.~of Physics, 
University of Mississippi--Oxford, University, MS 38677, USA} 

\address{$^2$\ Brookhaven National Laboratory, Upton, NY 11973, USA} 

\address{$^3$\ Dept.~of Physics, 
University of California, Los Angeles, CA 90095, USA} 

\begin{abstract}
A 4600 Hz fast ramping synchrotron is explored as an economical way of
accelerating muons from 4 to 20 GeV/c for a neutrino factory. Eddy current
losses are minimized by the low machine duty cycle plus thin grain
oriented silicon steel laminations and thin copper wires. Combined function
magnets with high gradients alternating within single magnets form
the lattice we describe. Muon survival is 83\%.

\end{abstract}


\vspace*{-8mm}
\submitto{J. Phys. {\bf G}}


\section{Introduction}

Traditionally ramping synchrotrons have provided economical particle
acceleration. Here we explore a very fast ramping muon synchrotron for a
neutrino factory \cite{factory}. The accelerated muons would be stored in a
racetrack to produce neutrino beams as they decay 
($\mu^- \to e^- \, {\overline{\nu}}_e \, \nu_{\mu}$ \, or \, 
$\mu^+ \to e^+ \, \nu_e \, {\overline{\nu}}_{\mu}$). Neutrino oscillations
\cite{oscillation} have been observed at experiments such as Homestake
\cite{homestake}, Super--Kamiokande \cite{superk}, and SNO \cite{SNO}. Further
exploration using a neutrino factory could reveal effects such as CP 
violation in the lepton sector
which could explain the matter--antimatter asymmetry of the universe.

This synchrotron must accelerate muons from 4 to 20 GeV/c
with moderate decay loss. Because synchrotron radiation goes as $m^4$,
muons radiate two billion times (\,$(105.7/.511)^4$\,) 
less power than electrons for any given ring
diameter and lepton energy. Magnet eddy current losses are minimized by the
low duty cycle of the machine plus thin iron laminations and 
copper conductors. Grain oriented silicon steel is used to provide a high
magnetic field with a high $\mu$ to minimize magnetic energy stored in the
return yoke. The magnetic energy stored in the gap is minimized by
reducing its size. Cool muons \cite{cool} with low beam emittance  
allow this. Stored energy
goes as $B^2/2\mu$. The voltage required to drive a magnet is equal to 
$-L \, di/dt$. Very high voltage is expensive. $di/dt$ must be large
because of the 2 $\mu$sec muon lifetime, so the main option for lowering
voltage is to shrink the volume of 
stored energy to reduce the
inductance, $L$. 

Acceleration to 4 GeV might feature fixed field dogbone arcs \cite{dogbone} to
minimize muon decay loss. Fast ramping synchrotrons \cite{dogbone, snowmass}
might also accelerate muons to higher energies for a $\mu^+ \, \mu^-$ collider
\cite{collider}.

\section{Lattices}

As a first step, we form arcs with sequences of combined function cells formed
within continuous long magnets, whose poles are alternately shaped to give
focusing gradients of each sign. An example of such a cell has been simulated
using SYNCH \cite{synch}. The example has gradients that alternate from
positive 20 T/m gradient (2.24 m long), to zero gradient (.4 m long) to
negative 20 T/m gradient (2.24 m) to zero gradient (0.4 m), etc. The
relatively short zero gradient section is included to approximate a real
smooth change in the gradients. Details are given in Table 1.

\begin{table}[!htb]
\begin{center}
\caption{Combined function magnet cell parameters. Five cells make up an
arc and 18 arcs form the ring.}
\vspace*{2mm}
\begin{tabular}{lcc}
\hline
Cell length &m&5.28\\
Combined Dipole length &m& 2.24\\
Combined Dipole B$_{\rm central}$ &T& 0.9\\
Combined Dipole Gradient &T/m& 20.2\\
Pure Dipole Length &m& 0.4\\
Pure Dipole B &T& 1.8\\
Momentum &GeV/c&20\\
\hline
Phase advance/cell&deg&72\\
beta max &m& 8.1\\
Dispersion max &m&0.392\\
\hline
Normalized Trans.~Acceptance & $\pi$ mm rad & 4 \\
\hline
\end{tabular}
\end{center}
\end{table}

It is proposed to use 5 such arc cells (possibly all in one magnet) to 
form an arc segment. These segments are alternated with straight 
sections containing RF. The phase advance through one arc segment is 5 x 
72 = 360 degrees. This being so, dispersion suppression between 
straights and arcs can be omitted. With no dispersion in the straight 
sections, the dispersion performs one full oscillation in each arc 
segment, returning to zero for the next straight as shown in Fig.~1. 
There will be 18 such arc segments and 18 straight sections, forming the 
18 superperiods in the ring.

\begin{figure}
\begin{flushleft}
\epsfxsize 144mm
\epsfbox{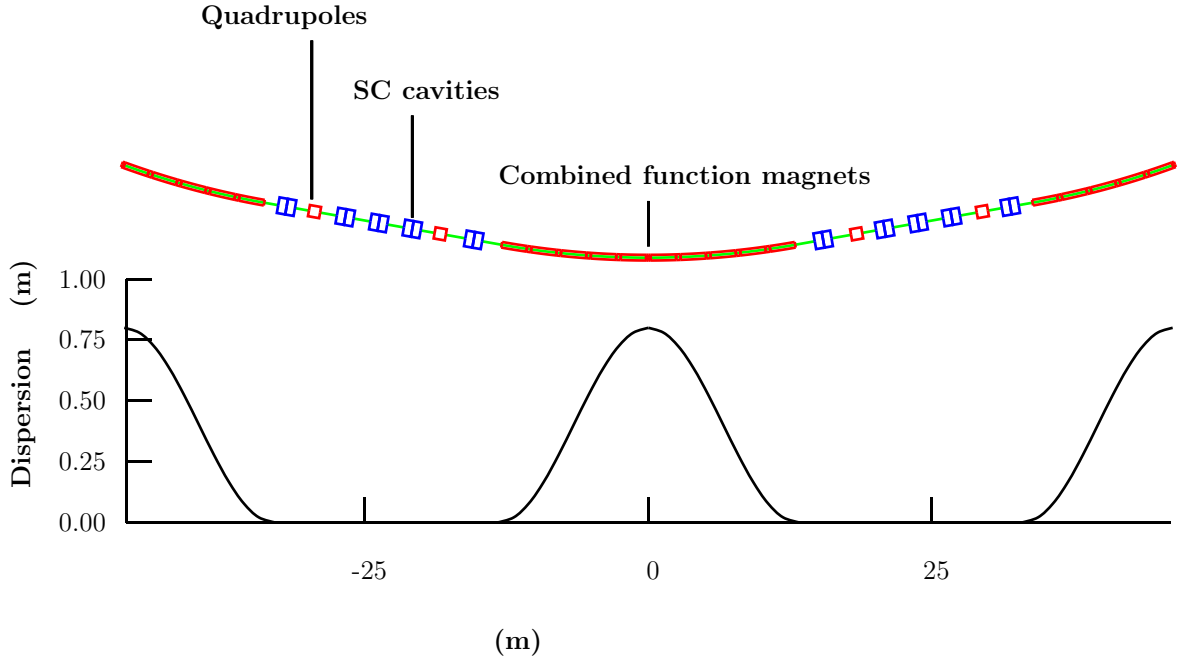}
\end{flushleft}
\vspace*{-6mm}
\caption{\label{label1}
Combined function magnets bend the muons in the arcs.  Superconducting
RF cavities accelerate muons in the straight sections. Two quadrupoles per
straight section provide focusing. The straight sections are dispersion free.}
\end{figure}

Straight sections (22 m) without dispersion are used for superconducting 
RF, and, in two longer straights (44 m), the injection and extraction. 
To assure sufficiently low magnetic fields at the cavities, 
relatively long field free regions are 
desirable. A straight consisting of two half cells 
would allow a central gap of 10 m between quadrupoles, and two 
smaller gaps at the ends. 
Details are given in Table 2.
Matching between the arcs and straights is not yet designed.
The total circumference of the ring including combined functions magnets
and straight sections adds up to 917 m 
($18 \times 26.5 \, + \, 16 \times 22 \, + \, 2 \times 44$). 

\begin{table}[!htb]
\begin{center}
\caption{Straight section lattice parameters. There are two quadrapoles per
straight.}
\vspace*{2mm}
\tabcolsep=2mm
\begin{tabular}{ccccccccc}
\hline
$\phi$ & L$_{\rm cell}$/2 & L$_{\rm quad}$ & dB/dx & a & $\beta_{\rm max}$ &
$\sigma_{\rm max}$ & B$_{\rm pole}$ & U$_{\rm mag}$/quad \\
77$^{\,0}$  & 11 m & 1 m & 7.54 T/m & 5.8 cm & 36.6 m &.0195 m & 0.44 T
& $\approx$ 3000 J \\ \hline
\end{tabular}
\end{center}
\end{table}

\section{Superconducting RF}

The RF must be distributed around the ring to 
avoid large differences between the beam momentum ( which increases in 
steps at each RF section) and the arc magnetic field (which is 
increasing continuously). RF parameters are shown in 
Table 3.  

  The amount of RF used is a tradeoff between cost and muon survival. Survival
is somewhat insensitive to the fraction of stored energy the beam
removes from the RF cavities, because the voltage drop is balanced by time
dilation. Here, only 8.2\% of RF energy is used.  One could,
in the spirit of Oliver Twist, ask for more. Using more of the RF
energy is particularly appealing with a smaller ring. Very cold muons
require less focusing and allow a smaller ring. 
If the muons take a few extra turns at the end
to accelerate, only a few extra will be lost. Also, 
extra acceleration
time at the end will translate into less voltage needed to ramp magnets.

\begin{table}[!htb]
\begin{center}
\caption{Superconducting RF parameters.}
\vspace*{2mm}
\tabcolsep=2mm
\begin{tabular}{lcc}
\hline
Frequency                 & 201                & MHz      \\
Gap                       & .75                & m       \\
Gradient                  & 15                 & MV/m     \\
Stored Energy             & 900                & Joules   \\
Muons per train           & $5 \times 10^{12}$ &          \\
Orbits (4 to 20 GeV/c)    & 12                 &          \\
No.~of RF Cavities        & 160                &          \\
RF Total                  & 1800               & MV       \\
$\Delta$U$_{\hbox{beam}}$ & 110                & Joules   \\
Energy Loading            & .082               &          \\
Voltage Drop              & .041               &          \\
Muon Acceleration Time    & 37                 & $\mu$sec \\
Muon Survival             & .83                &          \\ \hline            
\end{tabular}
\end{center}
\end{table}

\section{Combined Function Magnets}

The muons accelerate from 4 to 20 GeV.  If they are extracted at
95\% of full field they will be injected at 19\% of full field.
For acceleration with a plain sine wave, injection occurs at
11$^0$ and extraction occurs at 72$^0$.  So the phase must
change by 61$^0$ in 37 $\mu$sec.  Thus the sine wave goes through
360$^0$ in 218 $\mu$sec, which equals a frequency of 4600 Hz.
 
  Estimate the energy stored in each 26.5 m long combined function magnet.
The gap is about .14 m wide and has an average height of .06 m. Assume an
average field of 1.1 Tesla. The permeability constant, $\mu_0$, is $4\pi\times
10^{-7}$. $W = {B^2 / {2{\mu_0}}}[\hbox{Volume}] =$ 110\,000 Joules. Next
given one turn, an LC circuit capacitor, and a 4600 Hz frequency; estimate
current, voltage, inductance, and capacitance.

\begin{eqnarray}
\fl B = {{\mu_0\,NI}\over{h}}  \, \rightarrow\,
I = {{Bh}\over{\mu_0\,N}} = 52 \, \hbox{kA;} &
W = .5\,L\,I^2  \, \rightarrow\,  L = {2\,W / {I^2}} =
80\,\mu\hbox{H} \\
\fl f = {1\over{2\pi}}\sqrt{1\over{LC}}  \, \rightarrow\, 
C = {1\over{L\,(2\pi f)^2}} = 15\, \mu\hbox{F;} \  & \ 
W = .5\,C\,V^2  \, \rightarrow\,  V = \sqrt{2W / {C}} = 120\,\hbox{kV}
\end{eqnarray}
 
Separate coils might be put around each return yoke to halve the voltage as
illustrated in Fig.~2.  The stack of SCRs driving each coil might be center
tapped to halve the voltage again. Four equally spaced coil slots could be put
in each side yoke to cut the voltage by five, while leaving the pole faces
continuous. 6 kV is easier to insulate than 120 kV. It may be useful to shield
or chamfer \cite{school} magnet ends to avoid large eddy currents where the
field lines typically do not follow laminations. A DC offset power supply
\cite{white} could be useful. Neutrino horn power supplies 
look promising \cite{horn}.

Grain oriented silicon steel is chosen for the return yoke due to its
high permeability at high field at noted in Table 4.  This minimizes the
energy stored in the yoke which goes as $B^2/ 2 \mu$. 
The skin depth \cite{lorrain} 
of a
100 micron thick lamination is given by  

\begin{equation} 
\fl {\hbox{Skin Depth}} = \delta = \sqrt{\rho \, / \, \pi \, f \,\mu} = 
\sqrt{47\times{10^{-8}} \, / \, \pi \, 4600 \, 1000 \, \mu_0} = 
160 \mu{\hbox{m}}
\end{equation}

Take $\mu = 1000 \mu_0$ as a limit on magnetic saturation and hence energy
storage in the yoke. Next estimate the fraction of the inductance of the yoke
that remains after eddy currents shield the laminations \cite{lucent}.
The lamination thickness is $t$.

\begin{equation} 
\fl {\hbox{L/L}}_0 = (\delta/t) \, (\sinh(t/\delta) + \sin(t/\delta)) \,  / \,
(\cosh(t/\delta) + \cos(t/\delta)) = 0.995  
\end{equation}

So it appears that magnetic fields can penetrate 100 micron thick laminations 
at 4600 Hz.  If allowable, thicker 175 micron thick laminations would be 
half as costly and can achieve
a somewhat higher packing fraction.

\begin{table}[!htb]
\begin{center}
\caption{Approximate  permeabilities of soft magnetic materials.
The permeability is $B/\mu_0H$. Grain oriented silicon steel has a much
higher permeability parallel ($\parallel$) to its rolling direction than in
the perpendicular ($\perp$) direction \cite{armco,ferro}.}
\vspace*{2mm}
\renewcommand{\arraystretch}{1.05}
\tabcolsep=3mm
\begin{tabular}{lrrr} \hline 
Material                    &  1.0 Tesla & 1.5 Tesla & 1.8 Tesla \\ \hline 
1008 Steel                  &    3000 &   2000 &  200   \\
Grain Oriented ($\parallel$)&   40000 &  30000 & 3000   \\
Grain Oriented ($\perp$)    &    4000 & 1000   &        \\
NKK Super E-Core            &  20000  &    300 &   50   \\
Metglas 2605SA1             & 300000  &  10000 &    1   \\
\hline 
\end{tabular}
\end{center}  
\end{table}

Calculate the resistive energy loss in the copper coils, which over time
is equal to 1/2 the loss at the maximum current of 52\,000 amps.  The
1/2 comes from the integral of cosine squared. Table 5 gives the
resistivity of copper. Four 5\,cm square copper conductors each
5300\,cm long have a total power dissipation of 130 kilowatts/magnet.
Eighteen magnets give a total loss of 2340 kilowatts.
But the neutrino factory runs at 30 Hz.  Thirty half cycles 
of 109 $\mu$sec per second gives a duty factor of 300 and a total $I^2R$ loss
of 8000 watts.  Muons are orbited in opposite directions on alternate cycles. 
If this proves too cumbersome, the duty cycle factor could be lowered to 150.

\begin{equation}
\fl R = {5300 \ (1.8\,\mu\Omega\hbox{-cm})\over{(4) \, (5^2)}} = 95\,\mu\Omega;
\qquad
P = I^2R\int_0^{2\pi}\!\cos^2(\theta)\,d\theta = \hbox{130\,000 w/magnet}
\end{equation}

\begin{table}[!htb]
\begin{center}
\caption{Resistivity, magnetic saturation, and coercivity of conductors,
cooling tubes, and soft magnetic materials. The magnetic materials
include 
50, 100 \cite{arnold}, and 175 $\mu$m \cite{armco, ludlum} thick 
grain oriented silicon steel, 
NKK Super E-Core \cite{nkk}, and Metglas \cite{allied}.}
\vspace*{2mm}
\renewcommand{\arraystretch}{1.05}
\tabcolsep=1.3mm
\begin{tabular}{llcccl} \hline 
Material & Composition & $\rho$           & B$_{Max}$ & H$_c$ & Thicknesses  \\
         &             & $\mu\Omega$-cm   & Tesla     & Oe    & $\mu$m \\
                                                                         \hline
Copper             & Cu                      & 1.8       & --- & --- & ---   \\
Stainless 316L & 70\,Fe,\, 18\,Cr,\, 10\,Ni, 2\,Mo,\, .03\,C & 74 
& --- & --- & ---   \\
Titanium 6Al--4V & 90\,Ti,\, 6\,Al,\, 4\,V  & 171 & --- & --- & --- \\
1008 Steel         & 99\,Fe,\, .08\,C           & 12   & 2.09 &  0.8 & --- \\
Grain Oriented  
&  3\,Si,\, 97\,Fe           & 47                & 1.95 & .1 & 50,\,100,\,175 \\
NKK Super E-Core & 6.5\,Si,\, 93.5\,Fe       & 82    & 1.8  & .2 & 50,\,100  \\
Metglas 2605SA1
 & 81\,Fe,\, 14\,B,\, 3\,Si,\, 2\,C  & 135 & 1.6    & .03 & 30 \\
\hline 
\end{tabular}
\end{center}
\end{table}

Find the skin depth of copper at 4600 Hz to see if .25 mm (30 gauge)
wire is useable.
                                  
\begin{equation} 
\fl {\hbox{Skin Depth}} = \delta = \sqrt{\rho \, / \, \pi \, f \,\mu_0} = 
\sqrt{1.8\times{10^{-8}} \, / \, \pi \, 4600 \, \mu_0} = 
0.97 {\hbox{mm}}
\end{equation}

Now calculate the dissipation due to eddy currents in this .25 mm wide
conductor, which will
consist of transposed strands to reduce this loss \cite{sasaki, school}.
To get an idea, take the maximum B-field
during a cycle to be that generated by a 0.025m radius conductor carrying
26000 amps.
The eddy current loss in a rectangular conductor made of transposed square
wires .25 mm wide (sometimes called Litz wire \cite{mws}) 
with a perpendicular magnetic
field is as follows. The width of the wire is $w$ and
$B = {{\mu_0\,I}/{2\pi r}} = 0.2$ Tesla.

\begin{equation}
\fl P = \hbox{[Volume]}{{(2\pi\,f\,B\,w)^2}\over{24\rho}}
= [4 \ .05^2 \ 53]\, {{(2\pi \ 4600 \ .2 \ .00025)^2} \over
{(24)\,1.8\times{10^{-8}}}} = 2800 \
\hbox {kilowatts} 
\end{equation}

Multiply by 18 magnets and divide by a duty factor of 300
to get an eddy current loss in the copper of 170 kilowatts.
Stainless steel water cooling tubes will dissipate a similar amount
of power \cite{dogbone}. Alloy titanium cooling tubes would dissipate less.

Do the eddy current losses \cite{sasaki} in the 100 micron thick iron 
laminations. Take a quarter meter square area, a 26.5 meter length, and
an average field of 1.1 Tesla.

\begin{equation}
\fl {\hbox{P}
=  \hbox{[Vol]}{{(2\pi\,f\,B\,t)^2}\over{24\rho}}}
=  [(26.5) \, \, (.5^2)]\,
{{(2\pi \ 2600 \ 1.1 \ .0001)^2} \over
{(24)\,47\times{10^{-8}}}}
=  5900 \
\hbox {kw}
\end{equation}

Multiply by 18 magnets and divide by a duty factor of 300 to get an
eddy current loss in the iron laminations of 350 kilowatts or 700 watts/m
of magnet. 
So the iron will need some cooling. The ring only ramps 30 time per second, so
the $\int{\bf{H}}{\cdot}d\,{\bf{B}}$ hysteresis losses will be low, even
more so because of the low coercive force, H$_c$, of grain oriented silicon
steel.

\begin{figure}
\begin{flushleft}
\epsfxsize 136mm
\epsfbox{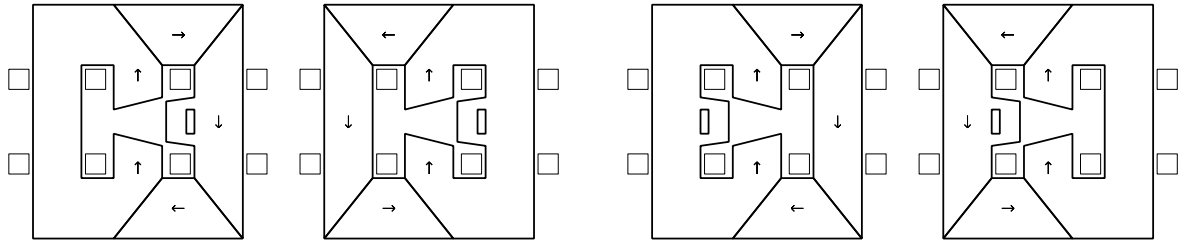}
\end{flushleft}
\vspace*{-6mm}
\caption{\label{label2}
Alternating gradient magnet laminations
with grain oriented silicon steel.  The
arrows show both the magnetic field direction and the grain direction of
the steel. If needed, four pieces might be used per layer as shown to fully 
exploit the high permeability and
low hysteresis in the grain direction 
\cite{school,sasaki,grain,transformer} noted in Table 4. 
The ``C" pieces provide rigidity.
Simpler solutions with one or two pieces per layer are under investigation.
The horizontal tab increases the gradient
by lowering the field to roughly zero on the wide side of the gap.
The four coils (\,\raisebox{.6ex}{\framebox[6pt]{\null}}\,) 
are wired in parallel.}
\end{figure}

\section{Conclusions}

The low duty cycle of the neutrino factory leads to reasonable eddy
current losses in a 4600 Hz ring. Muon survival is 83\%. The high
permeability of grain oriented silicon steel permits high fields with little
energy stored in the yoke. Gradients are switched within dipoles to minimize
eddy current losses in ends. Time dilation allows extra orbits with little
muon decay at the end of a cooling cycle. This 
allows one to use more of the stored RF energy. 
Much of the magnetic field in our lattice is used for focusing rather than
bending the muon beam. More muon cooling would lead to less
focusing, more bending, and an even smaller ring.

\section*{Acknowledgments}
This work was supported by the U.~S.~Dept.~of Energy and National Science 
Foundation.
Many thanks to K.~Bourkland, S.~Bracker, C.~Jensen,
S.~Kahn, H.~Pfeffer, G.~Rees, Y.~Zhao, and M.~Zisman for their help and 
suggestions.


\section*{References} 

\begin{thebibliography}{99}

\bibitem{factory}
Cline D and Neuffer D 1980 {\it AIP Conf. Proc.} {\bf 68} 846--7 \\ 
Neuffer D 1981 {\it IEEE Trans. Nucl. Sci.} {\bf 28} 2034--6 \\
Ayres D {\it et al} 1999 {\it Preprint}  physics/9911009 \\ 
Palmer R B, Johnson C and Keil E 2000 {\it Nucl.~Instrum.~Meth.} A
{\bf 451} 265--78 \\
Holtkamp N, Finley D A {\it et al} 2000 A feasibility study of a neutrino
source based on a muon storage ring {\it Preprint} FERMILAB-PUB-00-108-E \\  
Ozaki S, Palmer R B, Zisman M S, Gallardo J C {\it et al} 2001 Feasibility 
study II of a muon based neutrino source {\it Preprint} BNL-52623
http://www.cap.bnl.gov/mumu/studyii/
%


\bibitem{oscillation}
Barger V, Whisnant K and Phillips R J N 1980 {\it Phys.~Rev.~Lett.} {\bf 45}
2084--8 \\
Geer S 1998 {\it Phys.~Rev.} D {\bf 57} 6989--97 \\
Bilenky S M, Giunti C and Grimus W 1998 {\it Phys.~Rev.} D {\bf 58} 033001 \\
Albright C {\it et al} 2000 {\it Preprint} hep-ex/0008064 \\
Barger V, Geer S, Raja R and Whisnant K 2000
{\it Phys.~Rev.} D {\bf 62} 073002 \\
Barger V, Geer S, Raja R and Whisnant K 2000
{\it Phys.~Rev.} D {\bf 62} 013004 \\
Cervera A {\it et al} 2000 {\it Nucl.~Phys.} B {\bf 579} 17-55 \\
Romanino A 2000 {\it Nucl.~Phys.} B {\bf 574} 675--90 \\
De Rujula A, Gavela M B and Hernandez P 1999 {\it Nucl.~Phys.} B {\bf 547}
21--38 \\
Koike M and Sato J 2000 {\it Phys.~Rev.} D {\bf 61} 073012 \\
Kodama K {\it et al} (DONUT Collaboration) 2001 {\it Phys.~Lett.} 
{\bf B504} 218--24 
%

\bibitem{homestake}
Cleveland B T {\it et al} (Homestake Collaboration) 1998
{\it Astrophys. J.} {\bf 496} 505--26 \\ 
Davis R 1994 {\it Prog.~Part.~Nucl.~Phys.} {\bf 32} 13--32 \\ 
Davis R, Harmer D S and Hoffman K C 1968 {\it Phys.~Rev.~Lett.} {\bf 20} 
1205--9 \\
Davis R 1964 {\it Phys.~Rev.~Lett.} {\bf 12} 303--5 
%


\bibitem{superk}
Fukuda Y {\it et al} (Super-Kamiokande Collaboration) 1998
{\it Phys.~Rev.~Lett.} {\bf 81} 1562--7 \\
Fukuda S {\it et al} (Super-Kamiokande Collaboration) 2001
{\it Phys.~Rev.~Lett.} {\bf 86} 5651--5
%


\bibitem{SNO}
Ahmad Q R {\it et al} (SNO Collaboration) 2002 {\it Phys.~Rev.~Lett.} {\bf 89}
011301 \\ 
Ahmad Q R {\it et al} (SNO Collaboration) 2002 {\it Phys.~Rev.~Lett.} {\bf 89}
011302 \\
Ahmad Q R {\it et al} (SNO Collaboration) 2001 {\it Phys.~Rev.~Lett.} {\bf 87}
071301 \\
Boger J {\it et al} (SNO Collaboration) 2000 {\it Nucl.Instrum.Meth.} A 
{\bf 449} 172--207 \\
Chen H H 1985 {\it Phys.~Rev.~Lett.} {\bf 55} 1534--6  
%

\bibitem{cool}
Ado Y M and Balbekov V I 1971 {\it Sov.~Atom.~Energ.} {\bf 31} 731--6 \\ 
Skrinsky A N and Parkhomchuk V V 1981 {\it Sov.~J.~Part.~Nucl.} {\bf 12} 
223--47 \\
Neuffer D 1983 {\it Part.~Accel.} {\bf 14} 75--8 \\
Fernow R and Gallardo J 1995 {\it Phys.~Rev.} E {\bf 52} 1039--42 \\
Balbekov V I and Van Ginneken A 1998 {\it AIP Conf. Proc.} {\bf 441} 310--3 \\ 
Penn G and Wurtele J S 2000 {\it Phys.~Rev.~Lett.} {\bf 85} 764--7 \\
Wang C X and Kim K Y 2002 {\it Phys.~Rev.~Lett.} {\bf 88} 184801 \\
Alsharo'a M M {\it et al} 2002 {\it Preprint} hep-ex/0207031
%

\bibitem{dogbone} 
Summers D J 2001 Snowmass {\it Preprint} hep-ex/0208010 
%

\bibitem{snowmass} 
Summers D, Neuffer D, Shu Q S and Willen E 1997 PAC (Vancouver) {\it Preprint} 
physics/0109002 \\
Summers D J 1996 Snowmass {\it Preprint} physics/0108001 \\
Summers D J 1994 SESAPS (Newport News, VA) 
{\it Bull.~Am.~Phys.~Soc.} {\bf 39} 1818 
%


\bibitem{collider}
Neuffer D 1987 {\it AIP Conf. Proc.} {\bf 156} 201--8 \\ 
Cline D B 1994 {\it Nucl.~Instrum.~Meth.} {\bf A350} 24--6 \\
Neuffer D V 1994 {\it Nucl.~Instrum.~Meth.} {\bf A350} 27--35 \\
Barger V {\it et al} 1995 {\it Phys.~Rev.~Lett.} {\bf 75} 1462--5 \\ 
Palmer R {\it et al} 1996 {\it Nucl.~Phys.~Proc.~Suppl.} {\bf 51A} 61--84 \\ 
Raja R and Tollestrup A 1998 {\it Phys.~Rev.} D {\bf 58} 013005 \\   
Ankenbrandt C M {\it et al} 1999 {\it Phys.~Rev.~ST Accel.~Beams} {\bf 2} 
081001
%

\bibitem{synch}
Garren A A, Kenney A S, Courant E D and Syphers M J 1985 {\it Preprint}
FERMILAB-FN-420

\bibitem{horn}
Bourkland K, Roon K and Tinsley D (NuMI Collaboration) 2002 205 KA power 
supply for neutrino focusing horns {\it Preprint} FERMILAB-CONF-02-122-E

\bibitem{school}
Marks N 1994 Conventional Magnets -- I and II {\it CERN Accelerator School
Proceedings (University of Jyv\"{a}skyl\"{a})} CERN 94-01 Vol {\bf II} 
pp 867--911

\bibitem{white}
White M G, Shoemaker F C and O'Neill G K 1956 A 3 Bev high intensity
proton--synchrotron {\it CERN Symposium on High Energy Accelerators and Pion
Physics} CERN 56-25 Vol {\bf 1} pp 525--9 \\
Fox J A 1965 Resonant magnet network and power supply for the 4 GeV electron
synchrotron Nina {\it Proc. IEE} {\bf 112} 1107--26 \\
Westendorp W F 1945 {\it J.~Appl.~Phys.} {\bf 16} 657--60
%

\bibitem{lorrain}
Lorrain P, Corson D and Lorrain F 1988 {\it Electromagnetic Fields and Waves}
3rd edition (Freeman) pp 537--42

\bibitem{lucent}
Scott K L 1930 Variation of the inductance of coils due to the magnetic
shielding effect of eddy currents in the cores 
{\it Proc.~Inst.~Radio Eng.} {\bf 18} 1750--64
%

\bibitem{armco}
AK Steel (Butler, PA) http://www.aksteel.com/markets/electrical{\_}steels.asp

\bibitem{ferro} 
Bozorth R M 1951 {\it Ferromagnetism} (Van Nostrand) pp 90--1

\bibitem{arnold}
Arnold Engineering (Marengo, IL) http://www.grouparnold.com

\bibitem{ludlum}
Allegheny Ludlum (Pittsburgh, PA) http://www.alleghenyludlum.com

\bibitem{nkk}
NKK Corp (Tokyo) http://www.nkk.co.jp/en/products/steel/e-core/en/e-core.html

\bibitem{allied} 
Honeywell (Conway, SC) http://www.metglas.com

\bibitem{sasaki}
Sasaki H 1992 Magnets for fast--cycling synchrotrons {\it Talk at
International Conference on Synchrotron Radiation Sources (Indore)} KEK 91-216

\bibitem{mws}
MWS Wire Industries (Westlake Village, CA) http://www.mwswire.com/litzmain.htm

\bibitem{grain}
Schwandt P 1989 Comparison of realistic core losses in the booster ring
dipole magnets for grain--oriented and ordinary lamination steels
{\it Preprint} TRIUMPH--DN--89--K31

\bibitem{transformer}
Nakata T {\it et al} 1984 Influence of lamination
orientation and stacking on magnetic characteristics of grain oriented silicon
steel laminations {\it IEEE Trans. Magnetics} {\bf 20} 1774--9
%

\end{thebibliography}
\end{document}